\documentclass[11pt,onecolumn]{article}
\usepackage{amsmath,amsthm,amssymb,amsfonts,epsfig,graphicx}
\usepackage{color}

\usepackage[left=1.0in,top=1.0in,right=1.05in,bottom=0.5in,nohead,textheight=10in,footskip=0.3in]{geometry}

\newtheorem{theorem}{Theorem}

\newtheorem{definition}[theorem]{Definition}
\newtheorem{proposition}[theorem]{Proposition}

\begin{document}

\def\myparagraph#1{\vspace{2pt}\noindent{\bf #1~~}}



\def\calQ{{\cal Q}}
\def\Q{{Q}}

\def\DeltaCeil{{\lceil\Delta\rceil}}
\def\TwoDeltaCeil{{\lceil 2\Delta\rceil}}
\def\OnePointFiveDeltaCeil{{\lceil 3\Delta/2\rceil}}

\long\def\ignore#1{}
\def\myps[#1]#2{\includegraphics[#1]{#2}}
\def\etal{{\em et al.}}
\def\Bar#1{{\bar #1}}
\def\br(#1,#2){{\langle #1,#2 \rangle}}
\def\setZ[#1,#2]{{[ #1 .. #2 ]}}
\def\Pr{\mbox{\rm Pr}}
\def\REACHED{\mbox{\tt REACHED}}
\def\AdjustFlow{\mbox{\tt AdjustFlow}}
\def\GetNeighbors{\mbox{\tt GetNeighbors}}
\def\true{\mbox{\tt true}}
\def\false{\mbox{\tt false}}
\def\Process{\mbox{\tt Process}}
\def\ProcessLeft{\mbox{\tt ProcessLeft}}
\def\ProcessRight{\mbox{\tt ProcessRight}}
\def\Add{\mbox{\tt Add}}

\def\setof#1{{\left\{#1\right\}}}
\def\suchthat#1#2{\setof{\,#1\mid#2\,}} 
\def\event#1{\setof{#1}}
\def\q={\quad=\quad}
\def\qq={\qquad=\qquad}
\def\calA{{\cal A}}
\def\calC{{\cal C}}
\def\calD{{\cal D}}
\def\calE{{\cal E}}
\def\calG{{\cal G}}
\def\calL{{\cal L}}
\def\calN{{\cal N}}
\def\calP{{\cal P}}
\def\calR{{\cal R}}
\def\calS{{\cal S}}
\def\calT{{\cal T}}
\def\calU{{\cal U}}
\def\calV{{\cal V}}
\def\calO{{\cal O}}
\def\s{\footnotesize}
\def\calNG{{\cal N_G}}
\def\psfile[#1]#2{}
\def\psfilehere[#1]#2{}
\def\epsfw#1#2{\includegraphics[width=#1\hsize]{#2}}
\def\assign(#1,#2){\langle#1,#2\rangle}
\def\edge(#1,#2){(#1,#2)}
\def\VS{\calV^s}
\def\VT{\calV^t}
\def\slack(#1){\texttt{slack}({#1})}
\def\barslack(#1){\overline{\texttt{slack}}({#1})}
\def\NULL{\texttt{NULL}}
\def\PARENT{\texttt{PARENT}}
\def\GRANDPARENT{\texttt{GRANDPARENT}}
\def\TAIL{\texttt{TAIL}}
\def\HEADORIG{\texttt{HEAD$\_\:$ORIG}}
\def\TAILORIG{\texttt{TAIL$\_\:$ORIG}}
\def\HEAD{\texttt{HEAD}}
\def\CURRENTEDGE{\texttt{CURRENT$\!\_\:$EDGE}}

\def\unitvec(#1){{{\bf u}_{#1}}}
\def\uvec{{\bf u}}
\def\vvec{{\bf v}}
\def\Nvec{{\bf N}}

\newcommand{\bx}{\mbox{$x$}}
\newcommand{\by}{\mbox{\boldmath $y$}}
\newcommand{\bz}{\mbox{\boldmath $z$}}
\newcommand{\bu}{\mbox{\boldmath $u$}}
\newcommand{\bv}{\mbox{\boldmath $v$}}
\newcommand{\bw}{\mbox{\boldmath $w$}}
\newcommand{\bvarphi}{\mbox{\boldmath $\varphi$}}

\newcommand\myqed{{}}


\title{\Large\bf  \vspace{30pt} Minimizing a sum of submodular functions}
\author{Vladimir Kolmogorov \\ \normalsize University College London \\ {\normalsize\tt v.kolmogorov@cs.ucl.ac.uk}}
\date{}
\maketitle

\vspace{-6pt}
\begin{abstract}
We consider the problem of minimizing a function represented as a sum of submodular terms.
We assume each term allows an efficient computation of {\em exchange capacities}.
This holds, for example, for terms depending on a small number of variables,
or for certain cardinality-dependent terms.

A naive application of submodular minimization algorithms would not
exploit the existence of specialized exchange capacity subroutines for individual terms.
To overcome this,
we cast the problem as a {\em submodular flow} (SF) problem in an auxiliary graph,
and show that applying most existing SF algorithms would rely only on these subroutines.

We then explore in more detail Iwata's capacity scaling approach for submodular flows~\cite{Iwata:97}.
In particular, we show how to improve its complexity in the case when the function contains cardinality-dependent terms.
\end{abstract}

\section{Introduction}
\vspace{-6pt}
In this paper we consider the problem of minimizing an objective function of the following form:
\begin{equation}
f(S)=\sum_{Q\in\calQ^\circ} f_{Q}(S\cap Q)\qquad\quad \forall S\subseteq V
\label{eq:E}
\end{equation}
Here $V$ is a set of nodes, $\calQ^\circ\subseteq 2^V$ is a set of subsets of $V$,
and $f_Q:2^Q\rightarrow\mathbb R$ are submodular functions.

Function $f$ is itself submodular, and thus can be minimized in polynomial time.
The current fastest strongly
polynomial algorithms are those of Orlin~\cite{Orlin:MP09} and Iwata-Orlin~\cite{Iwata:SODA09},
which take time $O(n^5 EO + n^6 )$,
where $n=|V|$ and $EO$ is the time to run the value oracle for $f(S)$. The fastest weakly polynomial algorithms are those
of Iwata~\cite{Iwata:03} and Iwata-Orlin~\cite{Iwata:SODA09} which run in time $\tilde O(n^4 EO + n^5)$.

However, applying a general-purpose submodular minimization algorithm may not
be the most efficient technique, since it does not exploit the special structure of $f$.
It is often the case that terms $f_Q$ have a special form that allow an efficient
computation of exchange capacities, which are defined in the next section. Roughly speaking,
this means that we can efficiently minimize function $f_Q(S)-z(S)$
for any vector $z\in\mathbb R^Q$. (As usual, $z(S)$ denotes $\sum_{i\in S}z_i$.)
The main goal of this paper is to develop an algorithm that can exploit
the existence of specialized exchange capacities subroutines.

To achieve this goal, we use the framework of {\em submodular flows} (SF) introduced by Edmonds
and Giles~\cite{EdmondsGiles:77}. We show that the problem of minimizing $f$
can be cast as a particular SF instance in an auxiliary graph, so that
 computing exchange capacities for the new problem
is equivalent to computing exchange capacities for individual terms $f_Q$. 
Most existing algorithms for submodular flows rely on the exchange capacity oracle,
which gives the desired result.

We then present a capacity scaling technique for solving the problem.
Its complexity is $O((n+\sum_Q\alpha_Q)(n+\sum_Q\beta_Q)\log U)$
where $U$ is an upper bound on function values and $\alpha_Q$, $\beta_Q$ depend on the type of term $f_Q$:
\begin{itemize}
\item[(a)] If $|Q|=2$ then $(\alpha_Q,\beta_Q)=(1,1)$.
\item[(b)] If $f_Q(S)=g(|S|)$ then $(\alpha_Q,\beta_Q)=(|Q|,|Q|)$. Note, $g(\cdot)$ must be concave.
\item[(c)] If $f_Q(S)=g(|S\cap Q'|,|S\cap Q''|)$ where $Q',Q''$ are disjoint subsets of $Q$ then $(\alpha_Q,\beta_Q)=(|Q|^2,|Q|)$.
\item[(d)] For any other term $f_Q$ we have $(\alpha_Q,\beta_Q)=(|Q|^2,|Q|^2+|Q|\cdot h_Q)$
where $h_Q$ is the time of the exchange capacity oracle for the (scaled version of) $f_Q$.
\end{itemize}
\noindent In (b) and (c) we assume that function $g$ can be evaluated in $O(1)$ time.
For cases (c) and (d) we use the scaling technique of Iwata~\cite{Iwata:97}.

\myparagraph{Applications}
Functions with terms of the form (a)-(c) have recently appeared in computer vision
applications. Terms (a) and (b) were used for the {\em image segmentation} problem~\cite{Kohli:CVPR08,Vicente:ICCV09},
while terms (a) and (c) were used for {\em co-segmenting} two images containing a similar object~\cite{Hochbaum:ICCV09}.
(The latter work used terms of the form $f_Q(S)=-c\cdot |S\cap Q'|\cdot|S\cap Q''|$ with $c\ge 0$.)

Note, objective functions used in computer vision very often have form~\eqref{eq:E}
where $|Q|$ is quite small (2,3,$\ldots$). Terms $f_Q$ encode interactions between
neighboring pixels. Currently, researchers restrict themselves to functions
that can be reduced to a minimum $s$-$t$ cut problem (see discussion in~\cite{Zivny:DAM09}),
since minimizing general submodular functions is too expensive in practice.
Our work may remove such restriction.

\myparagraph{Related work} The problem of minimizing functions of the form~\eqref{eq:E}
was studied by Cooper~\cite{Cooper:08}, who formulated a linear program with an exponential number of constraints
and showed that its optimal value coincides with the minimum of $f$.
The formulation that we will use closely resembles that in~\cite{Cooper:08}.
Note, however, that the question of how to solve this LP efficiently was not addressed in \cite{Cooper:08},
and a connection to the submodular flow problem was not given.

It is known that in certain cases
the problem can be reduced to a minimum $s$-$t$ cut problem in a graph with auxiliary nodes.
Billionnet and Minoux~\cite{Billionnet:85} showed that this can be done for functions with cubic terms,
i.e.\ when $|Q|\le 3$ for all terms $f_Q$. Reductions for certain subclasses with
higher order terms were given by Freedman and Drineas~\cite{Freedman:CVPR05},
Kohli et al.~\cite{Kohli:CVPR08} and \v{Z}ivn\'{y} and Jeavons~\cite{Zivny:09}.
The resulting maxflow problem could be solved e.g. in $O(\min(\hat n^{2/3}, \hat m^{1/ 2}) \hat m \log(\hat n^2/\hat m) \log \hat U)$
time by the algorithm of Goldberg and Rao~\cite{Goldberg:ACM98}, 
where $\hat n$, $\hat m$ are the number of nodes and edges in the constructed graph and $\hat U$ is a bound on edge 
capacities.

On the negative side, \v{Z}ivn\'{y} et al.~\cite{Zivny:DAM09} proved
that some submodular terms with $|Q|=4$ do not admit such a reduction.
Even if the reduction exists, it may result in a graph which would be prohibitively large in practice.
Consider, for example, terms of the form $f_Q(S)=g(|S|)$
where $g$ is concave. The reduction of Kohli et al.~\cite{Kohli:CVPR08} 
adds $b$ extra nodes and $b|Q|$ extra edges for each term $f_Q$,
where $b$ is the number of breakpoints of the piecewise-linear concave function $g$.
If $g$ is strictly concave (as in the application of~\cite{Vicente:ICCV09})
then $b=|\Q|-1$, so there would be $O(|\Q|^2)$ edges.
In contrast, our technique uses only $O(|\Q|)$ memory.
The same holds for the function $f_Q(S)=-c\cdot |S\cap Q'|\cdot|S\cap Q''|$ used in~\cite{Hochbaum:ICCV09}.

Fujishige and Iwata~\cite{Fujishige:concave} considered functions of the form $f(S)+g(|S|)$
on a distributive lattice
where $f$ is submodular and $g$ is concave. It was shown that the problem
is equivalent to a {\em parametric} problem: minimize function of form $f(S)+c_\lambda(S)$
for all values of $\lambda$, where $\{c_\lambda\}_\lambda$ is a certain family of non-increasing vectors
in $\mathbb R^V$.

\vspace{-13pt}
\section{Problem formulation}
\vspace{-7pt}
Let $\calQ$ be the set obtained from $\calQ^\circ$ by removing all singleton subsets of the form $\{i\}$, $i\in V$.
Thus, $|Q|\ge 2$ for all $Q\in\calQ$.
Without loss of generality we assume that function $f$ is given by
\begin{equation}
f(S)=\sum_{i\in S} c_{it} + \sum_{i\in V-S} c_{si} + \sum_{Q\in\calQ} f_Q(S\cap Q)
\label{eq:energy}
\end{equation}
where $c_{it},c_{si}$ are non-negative numbers 
and each term $f_Q$ satisfies the following condition:\footnote{
If term $f_Q$ with $f_Q(\varnothing)=0$ does not satisfy~\eqref{eq:QAssumption} then
we can replace it with the sum $\varphi_Q(S)+\bar f(S)$
where $\bar f(S)=f(S)-\varphi_Q(S)$ and $\varphi_Q$ is a vector in the base polyhedron of $f_Q$, which can easy be computed by a greedy algorithm of Edmonds~\cite{Edmonds:70}.
}
\begin{equation}
\min_{S\subseteq 2^{\Q}} f_{\Q}(S)  =  f_{\Q}({\varnothing})=f_{\Q}(Q) = 0
\label{eq:QAssumption}
\end{equation}

\myparagraph{Base polyhedron and exchange capacities} The base polyhedron~\cite{Edmonds:70} of $f_Q$ is defined as
\begin{equation}
B(f_Q) = \{ \varphi_{Q}\in \mathbb R^Q \: | \: \varphi_Q(S)\le f_Q(S) \quad \forall S \subseteq Q, \;\; \varphi_Q(Q)=f_Q(Q)=0  \}
\label{eq:base'}
\end{equation}
Given a vector $\varphi_Q\in B(f_Q)$ and distinct nodes $i,j\in Q$, the exchange capacity $\bar c_{Qij}$ is the maximum value of $\epsilon\ge 0$
such that the operation $\varphi_{Qi}:=\varphi_{Qi}+\epsilon$, $\varphi_{Qj}:=\varphi_{Qj}-\epsilon$ keeps $\varphi_Q$ in $B(f_Q)$.
Clearly,
\begin{equation}
\bar c_{Qij}=\min_{S\subseteq Q} \{ \bar f_Q(S) \:|\: i\in S\subseteq Q-\{j\} \}\;\;\;,\quad \bar f_Q(S)=f_Q(S)-\varphi_Q(S)
\label{eq:cQij}
\end{equation}
Computing $\bar c_{Qij}$ is equivalent to minimizing a submodular function. 
This can be done in polynomial time by a number of general-purpose submodular minimization algorithms.
Furthermore, for many choices of $f_Q$ there exist more efficient specialized techniques. 

A remark on notation: in this paper we always use  ``bar'' ($\bar c_{Qij}$, $\bar f_Q$, $\ldots$)
to indicate ``residual'' values, i.e.\ values that take into account current flow.

\myparagraph{Maximum flow formulation}
Let us construct a directed capacitated graph $G=(N,A,c)$ as follows.
The set of nodes will be $N=\{s,t\}\cup V \cup_{Q\in\calQ} Q^\star$
where $s, t$ are the source and the sink and $Q^\star=\{Qi\:|\:i\in Q\}$ is a unique copy of $Q$.
Here $Qi$ is a shorthand notation for the pair $(Q,i)$.
The set of arcs will be
\begin{eqnarray*}
A&=&\{(i,Qi),(Qi,i)\:|\:i\in V,Qi\in N\} \bigcup \;  \{(s,i),(i,s),(i,t),(t,i)\:|\:i\in V\}
\end{eqnarray*}
Arc capacities $c_{si}$, $c_{it}$ are the same as in~\eqref{eq:energy}. Arcs to the source and from the sink
have zero capacity ($c_{is}=c_{ti}=0$), and all ``internal'' arcs have infinite capacity
($c_{i,Qi}=c_{Qi,i}=+\infty$).

A {\em flow} $\varphi$ is a vector in $\mathbb R^A$. For a subset $Q\in\calQ$ we denote $\varphi_Q\in\mathbb R^Q$
to be the vector with components $\varphi_{Qi}=\varphi_{i,Qi}$.
We also denote $value(\varphi)=\sum_{i\in V} \varphi_{si}$ to be the amount of flow sent from the source.
We will consider the following maximum flow problem:
\begin{subequations}\label{eq:maxflow}
\begin{eqnarray}
\max & & value(\varphi)\qquad\quad \mbox{s.t.} \\
 & & \varphi_{uv} = -\varphi_{vu}        \hspace{29pt} \forall (u,v) \in A \label{eq:antisymmetry} \mbox{\hspace{20pt}(antisymmetry)~~} \\
 & & \varphi_a \le c_a                   \hspace{49pt} \forall a\in A \label{eq:capacity}          \mbox{\hspace{37 pt}(capacity constraints)} \\
 & & \sum_{(u,i)\in A}\varphi_{ui} = 0   \hspace{21pt} \forall i \in V \label{eq:conservation}     \mbox{\hspace{38pt}(flow conservation for $V$)~~} \\
 & & \varphi_Q \in B(f_Q)                \hspace{31pt} \forall Q\in \calQ \label{eq:base}          \mbox{\hspace{32pt}(base polyhedron constraints)}  
\end{eqnarray}
\end{subequations}
Note, if $\varphi$ is feasible then we also have
$value(\varphi)\!=\!\!\sum_{i\in V}\varphi_{it}$ since 
$\sum_{i\in V}\varphi_{si}-\sum_{i\in V}\varphi_{it}
=\sum_{i\in V}[\varphi_{si}+\varphi_{ti}]
=-\sum_{i\in V}\sum_{Qi\in N}\varphi_{Qi,i}
=\sum_{Q\in\calQ}\sum_{i\in Q}\varphi_{i,Qi}=0$.

The linear program~\eqref{eq:maxflow} is very similar to that in~\cite{Cooper:08},
with some minor differences; for example, the ``balance'' constraint
$\varphi_Q(Q)=0$ is not present in~\cite{Cooper:08}.

The rest of the paper is organized as follows.
Section~\ref{sec:SF} gives a reduction of problem~\eqref{eq:maxflow} to a submodular flow problem, 
which leads to a number of algorithms for solving~\eqref{eq:maxflow}.
Section~\ref{sec:augmentation} describes a pseudo-polynomial
augmenting path algorithm, which is a specialization of the standard augmenting path algorithm
for submodular flows. By analyzing the algorithm we will prove that the maximum of~\eqref{eq:maxflow} coincides
with the minimum of $f$. Section~\ref{sec:scaling} presents a scaling version of the augmenting
path algorithm, while section~\ref{sec:implementation} discusses some implementational issues
and states the complexity of the algorithm.

The reader may choose to skip the next section; familiarity with the submodular flow problem
will not be necessary for understanding the augmenting path algorithm. 


\section{Reduction to a submodular flow problem}\label{sec:SF}
We will consider a directed capacitated graph $G'=(N,A',c)$
where $A'=A\cup\{(s,t),(t,s)\}$ and the capacities of the new arcs
are $c_{ts}=+\infty$, $c_{st}=0$. 
If $\varphi\in\mathbb R^{A'}$ is a flow in $G'$ and $u$ is a node in $N$ then $\partial \varphi(u)=\sum_{(v,u)\in A'} \varphi_{vu}$
will denote the amount of flow that comes into $u$. 

Let us recall a definition of a submodular flow problem for a graph $G'$~\cite{EdmondsGiles:77,Fujishige:IEICE00}. 
Assume that each arc $a\in A'$ has a cost $d_a$,
and let $g:2^N\rightarrow\mathbb R$ be a submodular function with $g(\varnothing)=g(N)=0$.
Then the problem is defined as
\begin{subequations}\label{eq:SF}
\begin{eqnarray}
\max  & &  \sum\nolimits_{a\in A'}  d_a \varphi_a \label{eq:SF:obj} \quad\qquad\mbox{s.t.}\\
 & & \varphi_{uv}=-\varphi_{vu}  \hspace{10pt} \forall (u,v)\in A' \label{eq:SF:antisymmetry} \\
                  & & \varphi_a\le c_a  \hspace{10pt} \forall a\in A' \label{eq:SF:capacity} \\
& & \partial \varphi \in B(g) 
\end{eqnarray}
\end{subequations}
where $B(g)$ is the base polyhedron of $g$:
\begin{equation}
B(g) = \{ z\in \mathbb R^N \: | \: z(X)\le g(X) \quad \forall X \subseteq N, \;\; z(N)=0  \}
\end{equation}

In order to simulate problem~\eqref{eq:maxflow}, we set arc costs as follows:
$d_{ts}=1$ and $d_a=0$ for all other arcs $a$.
Function $g$ is defined by
$$
g(X) = \sum_{\Q\in\calQ} f_Q(X^Q) 
$$
where we introduced notation $X^Q=\{i\in Q\:|\: Qi\in X\}$.

\begin{proposition}
Problems~\eqref{eq:maxflow} and~\eqref{eq:SF} are equivalent.
\end{proposition}
\begin{proof} 
Suppose that $\varphi\in\mathbb R^A$ is a feasible flow for problem~\eqref{eq:maxflow}. Let us extend it
to a flow in $G'$ by setting $\varphi_{ts}=value(\varphi)$, $\varphi_{st}=-value(\varphi)$.
Clearly, conditions~\eqref{eq:SF:antisymmetry} and~\eqref{eq:SF:capacity} are satisfied.
It is also easy to check that $z=\partial \varphi\in B(g)$. Indeed, we have $z_i=0$ for $i\in V\cup\{s,t\}$
and $z_{Qi}=\varphi_{Qi}$ for $Qi\in N$. Conditions $\varphi_{Qi}\in B(f_Q)$ then imply that $z(N)=0$ and for any $X\subseteq N$ there holds 
$z(X)=\sum_{Q\in\calQ} \varphi_Q(X^Q)\le \sum_{Q\in\calQ} f_Q(X^Q)=g(X)$.
Thus, $\varphi$ is a feasible flow for problem~\eqref{eq:SF}. Furthermore, the values of objective functions of~\eqref{eq:maxflow} and~\eqref{eq:SF} coincide.

Conversely, suppose that $\varphi\in\mathbb R^{A'}$ is a feasible flow for problem~\eqref{eq:SF};
let us show that its restriction to $A$ is feasible for problem~\eqref{eq:maxflow}.
Conditions~\eqref{eq:antisymmetry} and~\eqref{eq:capacity} follow from~\eqref{eq:SF:antisymmetry} and~\eqref{eq:SF:capacity}.
Denote $z=\partial \varphi$.
If $X$ is a subset of $N$ with $g(X)=g(N-X)=0$ then $z\in B(g)$ implies
$z(X)\le g(X)=0$ and $-z(X)=z(N-X)\le g(N-X)=0$, so $z(X)=0$.
Applying this fact for subset $X=\{i\}$ yields~\eqref{eq:conservation},
and applying this fact for subset $X=Q^\star$ yields constraint $\varphi_Q(Q)=0$, which is a part of~\eqref{eq:base}.
Finally, if $S\subseteq Q$ then $\varphi_{Qi}(S)=z(S^\star)\le g(S^\star)=f_Q(S)$ 
where we denoted $S^\star=\{Qi\:|\: i\in S\}$. Thus, $\varphi_Q\in B(f_Q)$. \myqed
\end{proof}

\myparagraph{Exchange capacities}
Most submodular flow algorithms rely on the following operation: given a feasible flow $\varphi\in\mathbb R^{A'}$
with $z=\partial\varphi\in B(g)$ and distinct nodes $u,v\in N$, compute the exchange capacity 
$\bar c_{uv}=\min_X \{ \bar g(X) \: | \: u\in X\subseteq N-\{v\}\}$ where $\bar g(X)=g(X)-z(X)$.
The proposition below shows that computing these capacities is equivalent to computing exchange capacities $\bar c_{Qij}$
for individual terms $f_Q$ with respect to flow $\varphi$ (given by eq.~\eqref{eq:cQij}).
\begin{proposition}
$\bar c_{uv}=\bar c_{Qij}$ if $(u,v)=(Qi,Qj)$ and $\bar c_{uv}=0$ otherwise.
\end{proposition}
\begin{proof}
As shown above, $z_i=0$ for $i\in V\cup\{s,t\}$, therefore
$z(X)=\sum_{Q\in\calQ}\varphi_{Qi}(X^Q)$ for all subsets $X\subseteq N$. This implies that
\begin{equation}
\bar g(X)=\sum_{Q\in\calQ} \bar f_Q(X^Q)
\end{equation}
The fact that $\varphi_Q\in B(f_Q)$ also implies $\min_{S\subseteq Q}\bar f_Q(S)=\bar f_Q(\varnothing)=\bar f_Q(Q)=0$ for all $Q\in\calQ$.
Therefore, if $(u,v)=(Qi,Qj)$ then the minimization problem $\min_X \{ \bar g(X) \: | \: u\in X\subseteq N-\{v\}\}$
has a minimizer $X\subseteq Q^\star$, and thus 
$\bar c_{uv}=\min_X \{ \bar g(X) \: | \: u\in X\subseteq Q^\star-\{v\}\}=\bar c_{Qij}$. Now suppose that $(u,v)\ne (Qi,Qj)$.
Let $U\subset N$ be the ``completion'' of $u$: $U=\{u\}$ if $u\in V\cup\{s,t\}$ and $U=Q^\star$ if $u=Qi$.
There holds $v\notin U$ since we assumed that $(u,v)\ne (Qi,Qj)$ and $u,v$ are distinct.
We have $\bar g(U)=0$, and thus $\bar c_{uv}=0$.
\myqed
\end{proof}

Problem~\eqref{eq:SF} is actually a {\em maximum submodular flow} problem,
which is a special case of the more general {\em minimum cost submodular flow} problem
(see survey~\cite{Fujishige:IEICE00}).
The former problem can be solved in time $O(|N|^3 h)$ by a push-relabel method 
of Fujishige and Zhang~\cite{FujishigeZhang:92}, where $h$ is the time of the exchange capacity oracle
(see also~\cite{Iwata:IPL00}, section 3.1).
Clearly, for certain functions $f$ this complexity can be better than
bounds $O(n^5 EO + n^6 )$ and $\tilde O(n^4 EO + n^5 )$ for submodular function minimization.

In our case $h$ is the maximum time of oracles over individual terms.
This appears to be a rather crude way of estimating the complexity, as it does not take into
account the structure of individual terms. We conjecture that a more careful analysis
of the algorithm can give a bound which better illustrates contributions of individual terms.
In the subsequent sections we will give an example of such a bound for 
a capacity scaling augmenting path algorithm applied to problem~\eqref{eq:maxflow}.


\section{Augmenting path algorithm}\label{sec:augmentation}
A shortest augmenting path algorithm for a problem equivalent to maximum submodular flows
was given by Fujishige~\cite{Fujishige:78}. We now describe its application to problem~\eqref{eq:maxflow},
and prove that the value of the maximum flow coincides with the minimum of $f$.
We will generalize the problem slightly: we assume that capacities $c_{is}$ and $c_{ti}$
are non-negative numbers which are not necessarily zero. (We will need this in the next section.)

Given a flow $\varphi$, the residual capacity for arc $a\in A$ is defined as $\bar c_a=c_a-\varphi_a$.
Similarly, we define ``residual functions'' $\bar f_Q$ by $\bar f_Q(S)=f_Q(S)-\varphi_Q(S)$ for $S\subseteq Q$.
It can be seen that if $\varphi$ satisfies antisymmetry
and conservation constraints~\eqref{eq:antisymmetry}, \eqref{eq:conservation} then for any $S\subseteq V$ there holds
\begin{equation}
f(S)=value(\varphi)+\sum_{i\in S}\bar c_{it}+\sum_{i\in V-S}\bar c_{si}+\sum_{Q\in\calQ}\bar f_Q(S\cap Q)
\label{eq:energy'}
\end{equation}
Indeed, subtracting~\eqref{eq:energy} from~\eqref{eq:energy'}
gives
$
\sum_{i\in V}\varphi_{si}-\sum_{i\in S} \varphi_{it} - \sum_{i\in V-S} \varphi_{si}
- \sum_{Q\in\calQ}\sum_{i\in S\cap Q} \varphi_{i,Qi}
=\sum_{i\in S}\left[\varphi_{si}+\varphi_{ti}+\sum_{Qi\in N}\varphi_{Qi,i}\right]=0
$.
All residual values for a feasible $\varphi$ are non-negative, so
equation~\eqref{eq:energy'} implies the weak duality relationship:
\begin{equation}
\max \{value(\varphi)\:|\:\varphi\mbox{~is feasible}\:\} \le \min \{f(S)\:|\:S\subseteq V\}
\end{equation}

Given a feasible flow $\varphi$, let $\bar A$ be the following set of arcs:
\begin{eqnarray}
&&\bar A=\{a\in A\:|\:\bar c_a > 0\}\bigcup\limits_{Q\in\calQ}\bar A_Q \;\;, \qquad
\bar A_Q=\{(Qi,Qj)\:|\: i,j\in Q,i\ne j,\bar c_{Qij}>0\}\label{eq:barA}
\end{eqnarray}
\begin{proposition}
If there is no path from $s$ to $t$ in $(N,\bar A)$ then
 the set $S=\{i\in V\:|\:i$ is reachable from $s$ in $(N,\bar A)\}$
satisfies $f(S)=value(\varphi)$, and therefore $\varphi$ is a maximum flow
and $S$ is a minimizer of $f$. 
\label{prop:noAugment}
\end{proposition}
\begin{proof}
It suffices to show that every term in the RHS of~\eqref{eq:energy'} (except maybe
for the first term $value(\varphi)$) is zero. If $i\in S$ then $\bar c_{it}=0$,
otherwise $t$ would be reachable from $s$.
If $i\in V-S$ then $\bar c_{si}=0$, otherwise $i$ would belong to $S$.
Consider the term for subset $Q\in\calQ$,
and denote $S'=S\cap Q$. For each pair of nodes $i\in S'$, $j\in Q-S'$
function $\bar f_Q$ must have a minimizer $S_{ij}$ with $i\in S_{ij}\subseteq Q-\{j\}$, otherwise we would have
$\bar c_{Qij}>0$ so node $j$ could be reached from $i$ via arcs
$(i,Qi),(Qi,Qj),(Qj,j)\in \bar A$ and thus $j$ would be in $S$.
The submodularity of $\bar f_Q$ implies that the set $\bigcup_{i\in S'}\bigcap_{j\in Q-S'}S_{ij}$
is a minimizer of $\bar f_Q$ as well. The latter set coincides with $S'=S\cap Q$, therefore $\bar f_Q(S\cap Q)=0$.
\myqed
\end{proof}

Now suppose that there exists a path $P$ from $s$ to $t$;
such a path is called an {\em augmenting path}.
Clearly, we can send some flow $\delta>0$
along the path\footnote{
Sending flow $\delta$ along arc $(u,v)\in A$ denotes
the operation
$\varphi_{uv}:=\varphi_{uv}+\delta$, 
$\varphi_{vu}:=\varphi_{vu}-\delta$.
Sending flow $\delta$ along arc $(Qi,Qj)\in \bar A_Q$ does not change $\varphi$.
} so that the flow would remain feasible and $value(\varphi)$ would increase by $\delta$.
This leads to
\begin{proposition}[Strong duality]
The value of the maximum flow in~\eqref{eq:maxflow} coincides with the minimum of $f$.
\end{proposition}
\begin{proof}
Let $\varphi$ be a maximum flow. There can be no augmenting path for $\varphi$, otherwise $\varphi$ would not be maximal.
The claim now follows from proposition~\ref{prop:noAugment}.
\myqed
\end{proof}
From now on, we assume that all capacities $c_{si}$, $c_{it}$
and values $f_Q(S)$ for $S\subseteq Q$ are integers bounded by constant $U$.
A maximum flow can then be computed in
pseudo-polynomial time by the following augmenting path algorithm: 
\begin{itemize}
\item[S0] Set $\varphi_a=0$ for all arcs $a$.
\item[S1] Construct set of arcs $\bar A$ as in~\eqref{eq:barA}.
\item[S2] Find a shortest path $P$ from $s$ to $t$ in $(N,\bar A)$; if no such $P$ exists, terminate.
\item[S3] Send 1 unit of flow along $P$ and go to step 1.
\end{itemize}
Note, it is well-known that for integer-valued submodular flow problems 
sending 1 unit of flow along a {\bf shortest} augmenting path preserves flow feasibility~\cite{Fujishige:78}.
In our case we can relax slightly the requirement
that $P$ is shortest; we only need $P$ to be {\em minimal}:

\begin{definition}
Let $P$ be a simple (i.e.\ node-disjoint) path in $(N,\bar A)$. We call $P$ {\em minimal}
(with respect to $(N,\bar A)$) if the following property holds:
if $(Qi,Qj)$, $(Qi',Qj')$ are two distinct arcs in the path (occurring in this order)
then $\bar A$ does not have arc $(Qi,Qj')$.
\end{definition}
Clearly, any shortest augmenting path from $s$ to $t$ is minimal.
In Appendix A we prove that sending one unit of flow from $s$ to $t$ along a minimal path
preserves flow feasibility.

It is not difficult to show that sets $\bar A_Q$ are transitive, i.e.\ $(i,j),(j,k)\in \bar A_Q$
implies $(i,k)\in \bar A_Q$ (see Appendix A).
Thus, if $P$ is minimal then $(Qi,Qj)\in P$ implies
that the previous arc in $P$ is $(i,Qi)$ and the next arc is $(Qj,j)$.
The operation of sending flow through these three arcs 
will be referred to as ``sending flow from $i$ to $j$ via $Q$''.


\begin{figure}[!t]
\centering{
\framebox[156mm][l]{
\begin{minipage}[l]{148mm}

\begin{itemize}
\item[S0] 
For each $Q\in\calQ$ set $\varphi_Q\!:=\!\AdjustFlow^\Delta_Q(\varphi_Q)$
to make sure that $\varphi_Q\in B(f^\Delta_Q)$. 
Adjust other flow components so that $\varphi$ satisfies
antisymmetry and flow conservation constraints:
\begin{itemize}
\item[$\bullet$] Set $\varphi_{Qi,i}:=-\varphi_{i,Qi}$ for all $Qi\in N$.
\item[$\bullet$] For each node $i\in V$ compute $\delta=\sum_{(u,i)\in A}\varphi_{ui}$;
if $\delta>0$, send $\delta$ units of flow back to the source via arc $(i,s)$,
otherwise send $-\delta$ units of flow from the sink via arc $(t,i)$.
\end{itemize}
\item[S1] Construct set of arcs $\bar A^{\Delta}$ as follows:
\begin{eqnarray}
\bar A^\Delta&=&\{(u,v)\in A\:|\:\bar c_{uv}\ge \DeltaCeil \} 
\;\bigcup_{Q\in \calQ} \bar A^\Delta_Q
\end{eqnarray}
\item[S2] Find minimal path $P$ in $(N,\bar A^\Delta)$; if no such path exists, terminate.
\item[S3] Send $\DeltaCeil$ units of flow along $P$ and go to step S1.
\end{itemize}

\end{minipage}
} 
} 
\caption[]{\it
{\bf $\Delta$-phase.} Definitions of function $f^\Delta_Q$, procedure $\AdjustFlow^\Delta_Q(\varphi)$ and set $\bar A^\Delta$ 
for different types of terms $f_Q$ are given in sections \ref{sec:pairwise}-\ref{sec:general}.
}
\label{fig:CapScaling}
\end{figure}

\section{Capacity scaling algorithm}\label{sec:scaling}
We now apply a scaling technique to get a weakly-polynomial algorithm. 
As usual, the algorithm works in phases. Each phase is associated with a number $\Delta=2^l$, $l=-1,0,1,2,\ldots$; we call it a $\Delta$-phase.
To initialize, we set $\Delta=2^{\lceil \log_2 U \rceil}$ and $\varphi_a=0$ for all arcs $a\in A$.
After completing the $\Delta$-phase we divide $\Delta$ by 2
and proceed to the next phase (or terminate, if $\Delta=1/2$).
The $\Delta$-phase is described in Figure~\ref{fig:CapScaling}.
This description uses the following yet undefined objects:
\begin{itemize}
\item $f^\Delta_Q$ is a submodular function. When $\Delta=\frac{1}{2}$, function $f^\Delta_Q$ coincides with $f_Q$.
\item $\AdjustFlow^\Delta_Q(\varphi_Q)$ is a procedure that outputs a vector in $B(f^\Delta_Q)$
whose components are integer multiples of $\DeltaCeil$.
\item $\bar A^\Delta_Q$ is a subset of arcs of the form $(Qi,Qj)$ where $i,j$ are distinct nodes in $Q$. 
Set $\bar A^\Delta_Q$ is transitive, i.e.\ $(Qi,Qj),(Qj,Qk)\in\bar A^\Delta_Q$ for distinct $i,j,k\in Q$
implies $(Qi,Qk)\in \bar A^\Delta_Q$.
When $\Delta=\frac{1}{2}$, set $\bar A^\Delta_Q$ coincides with the set $\bar A_Q$ defined in \eqref{eq:barA}.
\end{itemize}
Definitions of these three objects will depend on the type of term $f_Q$;
different cases are considered in sections \ref{sec:pairwise}-\ref{sec:general}. Set $\bar A^\Delta_Q$ will be defined in such a way
that each augmentation keeps flow $\varphi_Q$ in $B(f^\Delta_Q)$.

It is clear that each
$\Delta$-phase maintains the following invariants: (i) components of flow $\varphi$ are integer multiples of $\DeltaCeil$;
(ii) $\varphi$ is a feasible $\Delta$-flow, i.e.\ it satisfies
 antisymmetry \eqref{eq:antisymmetry}, capacity \eqref{eq:capacity}, flow conservation constraints \eqref{eq:conservation}, as well as base polyhedron constraints $\varphi_Q\in B(f^\Delta_Q)$.
(We assume that capacities $c_{is}$, $c_{ti}$ are infinite, so that sending flow to the source or from the sink
in step S0 is always feasible).
To estimate the complexity, we will use values $\alpha_Q$ (to be defined in sections \ref{sec:pairwise}-\ref{sec:general}) that satisfy
\begin{equation}
\bar f^\Delta_Q(S) + \sum_{i\in Q} |\varphi_{Qi}-\varphi^\circ_{Qi}| \le \alpha_Q\cdot \DeltaCeil
\label{eq:alpha}
\end{equation}
where $\varphi^\circ$ is the flow in the beginning of $\Delta$-phase,
$S$ is the set of nodes in $Q$ reachable from $s$ in the graph $(N,\bar A^{2\Delta})$
constructed with respect to flow $\varphi^\circ$,
$\varphi=\AdjustFlow^\Delta_Q(\varphi^\circ)$ and $\bar f^\Delta_Q(S)=f^\Delta_Q(S)-\varphi_Q(S)$. 
Values $\alpha_Q$ can be used for estimating the number of augmentations (a proof is given in Appendix B):
\begin{proposition}
Each $\Delta$-phase terminates after at most $2n+\sum_{Q\in\calQ}\alpha_Q$ 
augmentations, and so the whole algorithm performs $O((2n+\sum_{Q\in\calQ}\alpha_Q)\log U)$ augmentations.
\label{prop:alpha}
\end{proposition}
To complete the description of the algorithm, we need to provide
constructions for different types of terms $f_Q$. 
In sections \ref{sec:pairwise}-\ref{sec:general} below we 
consider three types: pairwise terms, cardinality-dependent terms and general terms.

\vspace{-4pt}
\subsection{Pairwise terms}\label{sec:pairwise}
First, we consider the case when $|Q|=2$, which occurs very frequently in applications (see e.g.~\cite{BK:PAMI04} for a survey of applications in computer vision).
We define $f^\Delta_Q=f_Q$ for all $\Delta$. This means that procedure $\AdjustFlow^\Delta_Q(\varphi_Q)$
can simply return $\varphi_Q$ - it is guaranteed to belong to $B(f^\Delta_Q)=B(f^{2\Delta}_Q)$.
Let $Q=\{i,j\}$. Constraint $\varphi_Q\in B(f^\Delta_Q)$ can be written as
\begin{equation}
\varphi_{Qi}  \le  f_Q(\{i\}) \qquad
\varphi_{Qj}  \le  f_Q(\{j\}) \qquad
\varphi_{Qi} + \varphi_{Qj}  =  0
\label{eq:polyhedron:Q0}
\end{equation}
The set of arcs $\bar A^\Delta_Q$ is constructed as follows:
we add arc $(i,j)$ if $\bar c_{Qij}=\bar f_Q(\{i\})\ge \DeltaCeil$, and arc $(j,i)$ if $\bar c_{Qji}=\bar f_Q(\{j\})\ge \DeltaCeil$.
Clearly, we can always push $\DeltaCeil$ units of flow through the added arcs - constraints~\eqref{eq:polyhedron:Q0}
will be preserved.

It is easy to see that we can take $\alpha_Q=2$.
Indeed, let $S$ be the set used in eq.~\eqref{eq:alpha}.
Since $\varphi=\varphi^\circ$, we need to show that $\bar f_Q(S)\le 2\DeltaCeil$.
If $S=\{i\}$ then $(Qi,Qj)\notin \bar A^{2\Delta}$, therefore $\bar f_Q(S)\le \TwoDeltaCeil-1\le 2\DeltaCeil$.
The case $S=\{j\}$ is similar. If $S$ is empty or equals $Q$ then $\bar f_Q(S)=0$.



\vspace{-4pt}
\subsection{Cardinality-dependent terms}\label{sec:cardinality}
Let us now assume that $f_Q(S)$ for $S\subseteq Q$ depends only on $|S|$.
Thus, $f_Q(S)=g(|S|)$ where $g$ is a concave function.
As above, we define $f^\Delta_Q=f_Q$
for all $\Delta$, and accordingly procedure $\AdjustFlow(\varphi_Q)$ simply returns $\varphi_Q$.
Below we describe how to construct set $\bar A^\Delta_Q$.

For integer numbers $a\le b$ let $\setZ[a,b]$
be the set of integers in $[a,b]$. We can assume that $g(\cdot)$ is defined only on $\setZ[0,m]$ where $m=|Q|$.
We denote $z=\varphi_Q$, so $z_i=\varphi_{i,Qi}=-\varphi_{Qi,i}$ for $i\in Q$.
For a vector $z\in\mathbb R^Q$ we also denote $(z^{1},\ldots,z^m)$ to be the sequence
of values of $z_i$ sorted in the non-increasing order.
Thus, $z^k$ is the $k$-th largest number among values $z_i$, $i\in Q$. For a node $i\in Q$ define
\begin{equation*}
L(i)=\min \{k\in\setZ[1,m] \:|\: z^k = z_i \}
\;,\quad
R(i)=\max \{k\in\setZ[1,m] \:|\: z^k = z_i \}
\end{equation*}

Let us define ``residual'' function $\bar g(\cdot)$ by $\bar g(k) = \min \{ \bar f_Q(S)\:|\:S\subseteq Q,|S|=k\}$
for $k\in\setZ[0,m]$.
We have\vspace{-5pt}
\begin{equation}
\bar g(k) 
= g(k) - \max \{ z(S)\:|\:S\subseteq Q,|S|=k\}
= g(k) - \sum_{k'=1}^k z^{k'}
\end{equation}
Clearly, constraint $z\in B(f_Q)$ is equivalent to the following conditions:
(i) function $\bar g(\cdot)$ is non-negative, i.e.\ $\bar g(k)\ge 0$ for all $k\in\setZ[0,m]$; (ii) $z(Q)=0$.

Recall that sending flow $\DeltaCeil$ from $i$ to $j$ via $Q$ denotes the following operation:
$
z_{i} := z_{i} + \DeltaCeil  , \; z_{j} := z_{j} - \DeltaCeil
$.
Next, we describe the effect of this operation on function $\bar g(\cdot)$.
Three cases are possible (we assume that we are in a $\Delta$-phase, so all
components of vector $z$ are integer multiples of $\DeltaCeil$):
\begin{itemize}
\item $z_i \le z_j - 2\DeltaCeil$.
The change in the sequence $(z^1,\ldots,z^m)$ is
$$(\ldots,0,-\DeltaCeil,0,\ldots,0,+\DeltaCeil,0,\ldots)$$ where $-\DeltaCeil$ is in the position $R(j)$
and $+\DeltaCeil$ is in the position $L(i)$.
Therefore, the effect of the operation is that all values $\bar g(k)$ for $k\in \setZ[R(j),L(i)-1]$
are increased by $\DeltaCeil$. 
\item $z_i = z_j - \DeltaCeil$. The values $z_i$ and $z_j$ are swapped, therefore the sequence
$(z^1,\ldots,z^m)$ and function $\bar g(\cdot)$ do not change.
\item $z_i \ge z_j$.
The change in the sequence $(z^1,\ldots,z^m)$ is
$$(\ldots,0,+\DeltaCeil,0,\ldots,0,-\DeltaCeil,0,\ldots)$$ where $+\DeltaCeil$ is in the position $L(i)$
and $-\DeltaCeil$ is in the position $R(j)$.
Therefore, all values $\bar g(k)$ for $k\in\setZ[L(i),R(j)-1]$
are decreased by $\DeltaCeil$. 
\end{itemize}

In the first two cases function $\bar g(\cdot)$ cannot become negative,
thus sending $\DeltaCeil$ units of flow from $i$ to $j$ via $Q$ is always possible if $z_i < z_j$.
Accordingly, we add arcs $(i,j)$ to $\bar A^\Delta_Q$ for all pairs of nodes $i,j\in Q$ with $z_i < z_j$.
If $z_i \ge z_j$ then we can send flow if and only if $\min_{k\in \setZ[L(i),R(j)-1]} \bar g(k)\ge \DeltaCeil$.
However, if we add all arcs that satisfy this constraint then sending $\DeltaCeil$ units of flow
through {\bf multiple} arcs of $Q$ along a minimal path could make some values $\bar g(k)$ negative.
To prevent this, we add to $\bar A^\Delta_Q$ those arcs $(i,j)$ with $z_i \ge z_j$
that satisfy the following constraint:
\begin{equation}
\min_{k\in \setZ[L(i),R(j)-1]} \bar g(k) \ge 3\Delta/2
\end{equation}
The proposition below shows the correctness of this construction, and gives a bound on $\alpha_Q$.
A proof is given in Appendix C.
\begin{proposition}
(a) Set $\bar A^\Delta_Q$ is transitive.
(b) Sending $\DeltaCeil$ units of flow through a minimal path $P$ in $(N,\bar A^\Delta)$
preserves constraint $\varphi_Q=z\in B(f_Q)$. (c) Eq.~\eqref{eq:alpha} is satisfied by $\alpha_Q=3(m-1)$.
\label{prop:cardinality}
\end{proposition}

\vspace{-11pt}
\subsection{General submodular terms}\label{sec:general}
For general terms we can use the technique of Iwata~\cite{Iwata:97}.  $f^\Delta_Q$ is defined as
\begin{equation}
f^\Delta_Q(S) = \Delta\cdot \lfloor f_Q(S)/\Delta \rfloor + \lfloor \Delta \rfloor \cdot b(S) \qquad \forall S\subseteq Q
\end{equation}
where $b(S)=|S|\cdot|Q-S|$. As shown in~\cite{Iwata:97}, this function is submodular. The set $\bar A^\Delta_Q$ includes
all arcs $(Qi,Qj)$ that have non-zero residual capacity with respect to function $\bar f^\Delta_Q(S)=f^\Delta_Q-\varphi_Q(S)$.
Clearly, values of $f^\Delta_Q(S)$ are integer multiples of $\DeltaCeil$, so results in section~\ref{sec:augmentation}
imply that pushing $\DeltaCeil$ of flow through a minimal path in $(N,\bar A^\Delta)$ preserves constraint $\varphi_Q\in B(f^\Delta_Q)$.

Procedure $\AdjustFlow^\Delta_Q(\varphi^\circ_Q)$ works as follows. First, define vector $\varphi'_Q$ by $\varphi'_{Qi}=\varphi^\circ_{Qi}-m\DeltaCeil$
where $m=|Q|$. Vector $\varphi'_Q$ belongs to submodular polyhedron 
\begin{equation}
P(f^\Delta_Q) = \{ \varphi_{Q}\in \mathbb R^Q \: | \: \varphi_Q(S)\le f^\Delta_Q(S) \quad \forall S \subseteq Q  \}
\end{equation}
Indeed, for any $S\subseteq Q$ we have $\varphi'_Q(S)=\varphi^\circ_Q(S)-m\DeltaCeil\cdot|S|\le f^{2\Delta}_Q(S)-m\DeltaCeil\cdot|S|
\le f^\Delta_Q(S)$.
Since $\varphi'_Q\in P(f^\Delta_Q)$, there exists vector $\varphi_Q\in B(f^\Delta_Q)$ with $\varphi'_Q\ge \varphi_Q$, which
can be found by a greedy algorithm starting from $\varphi'_Q$ \cite[Theorem 3.19]{Fujishige:91}. This $\varphi_Q$ is taken as the output of $\AdjustFlow^\Delta_Q(\varphi^\circ_Q)$.

It can be seen that $\alpha_Q=O(m^2)$. This follows from three facts: 
(1) $\bar f^{2\Delta}_Q(S)=0$ where $S$ is the set used in eq.~\eqref{eq:alpha}
and $\bar f^{2\Delta}_Q$ is the residual function with respect to flow $\varphi^\circ$; 
(2) $|f^{2\Delta}_Q(S)-f^{\Delta}_Q(S)|=O(m^2\DeltaCeil)$;
(3) $\sum_{i\in Q}|\varphi_i-\varphi^\circ_i|\le m^2\DeltaCeil +\sum_{i\in Q}|\varphi_{Qi}-\varphi'_{Qi}|=m^2\DeltaCeil + \varphi_{Q}(Q)-\varphi'_{Q}(Q)=2m^2\DeltaCeil$.

Note, procedure $\AdjustFlow^\Delta_Q(\varphi_Q)$ used in~\cite{Iwata:97} is slightly more complicated; in particular
it takes into account set $S$ used in~\eqref{eq:alpha}. However, both techniques lead to $\alpha_Q=O(m^2)$.

\vspace{-7pt}
\section{Efficient implementation}\label{sec:implementation}
\vspace{-2pt}
We now discuss how implement steps S1 and S2 of the algorithm, i.e.\ how to find a minimal augmenting
path. Set $\bar A^\Delta$ contains $O(n+\sum_Q |Q|^2)$ arcs, so a naive computation would take
$O(n+\sum_Q |Q|^2)$ time. However, this can easily be improved: it can be seen that an explicit
construction of $\bar A^\Delta$ is not required.

We will use a breadth-first search (BFS) for computing a shortest path from $s$ to $t$
in $(N,\bar A^\Delta)$. Each node $Qi\in N$ will have flag $\REACHED(Qi)$,
which is set to $\false$ at the beginning of BFS.
We assume that each term $f_Q$ supports operation $\GetNeighbors^\Delta_Q(Qi)$
for a node $Qi\in N$ with $\REACHED(Qi)=\false$. This operation is defined as follows:
\begin{itemize}
\item Compute $S=\{Qj\in N\:|\:(Qi,Qj)\in\bar A^\Delta_Q,\;\; \REACHED(Qj)=\false\}$.
\item Set $\REACHED(Qj):=\true$ for $Qj\in S\cup\{Qi\}$.
\item Return $S$ as a linked list.
\end{itemize}
Flags $\REACHED(Qi)$ will not be modified by any other operation
(except that they are reset to $\false$ at the beginning of BFS).

It is straightforward to implement the BFS procedure using operations \linebreak $\GetNeighbors^\Delta_Q(Qi)$.
The running time of one augmentation (steps S1-S3) will then be $O(n+\sum_{Q\in\calQ} \beta_Q)$
where $\beta_Q$ for a fixed $Q\in\calQ$ is the combined time taken by calls to $\GetNeighbors^\Delta_Q(Qi)$,
plus the time for sending flow through $Q$ in step S3 (which may update internal structures for $Q$). 
In Appendix D we show how to implement $\GetNeighbors^\Delta_Q(Qi)$ so that $\beta_Q=O(|Q|)$ in the following cases:
\begin{itemize}
\item $f_Q(S)=g(|S|)$ for $S\subseteq Q$.
\item $f_Q(S)=g(|S\cap Q'|,|S\cap Q''|)$ where $Q',Q''$ are disjoint subsets of $Q$.
\end{itemize}
The second case relies on the algorithm of Aggarwal et al.~\cite{Aggarwal:87} which
computes row minima of a totally monotone matrix in linear time.
For a general submodular term $f_Q$ a naive implementation
of $\GetNeighbors^\Delta_Q(Qi)$ would make $|Q|-1$ calls to the exchange capacity oracle for $f^\Delta_Q$,
giving $\beta_Q\,=\,O(|Q|^2 h_Q)$ where $h_Q$ is the oracle's complexity.
However, the set $\{(Qi,Qj) \: |  \: (Qi,Qj) \in \bar A^\Delta_Q\}$ can 
alternatively be obtained from the minimal minimizer in $\arg \min \{ \bar f^\Delta_Q(S) \: | \: i\in S \subseteq Q \}$.
It is natural to assume that computing such minimal minimizer also takes time $h_Q$.
Under this assumption $\beta_Q=O(|Q|^2+|Q|\cdot h_Q)$. Combined with proposition~\ref{prop:alpha},
this leads to the overall complexity stated in the introduction.


\section{Conclusions and future work}
In recent years there has been an increased interest in the computer vision community in using 
submodular functions of the form~\eqref{eq:E} with high-order terms~\cite{Kohli:CVPR08,Kohli:PAMI09,Vicente:ICCV09,Hochbaum:ICCV09,Delong:CVPR10}.
So far, researchers restricted themselves to functionals that can be transformed to pairwise terms by introducing auxiliary variables.
The main goal of this paper is to advocate a more direct approach which could extend the set of practically tractable functionals.

To our knowledge, our bound $O((n+\sum_Q\alpha_Q)(n+\sum_Q\beta_Q)\log U)$ is the first one for minimization problem~\eqref{eq:E}
that shows contributions of individual terms. It is quite likely, however, that it can be improved further.
Indeed, the capacity scaling algorithm of Iwata~\cite{Iwata:97} that we built on is not a state-of-the-art.
In the future we plan to investigate applications of alternative submodular flow algorithms, such
as the capacity scaling algorithm of Fleischer et al.~\cite{Fleischer:02} that improves on~\cite{Iwata:97},
or the push-relabel method of Fujishige and Zhang~\cite{FujishigeZhang:92}.


\section*{Appendix A: Minimal augmenting paths}
First, let us show the set $\bar A_Q$ defined in~\eqref{eq:barA} is transitive,
i.e.\ if $i,j,k$ are distinct nodes in $Q$ then $(Qi,Qj),(Qj,Qk)\in\bar A_Q$ implies $(Qi,Qk)\in\bar A_Q$.
Suppose not, then $\bar c_{Qik}=0$.
This means that $\bar f_Q(S)=0$ for some subset $S$ with $i\in S\subseteq Q-\{k\}$.
If $j\in S$ then $\bar c_{Qjk}=0$ and $(Qj,Qk)\notin \bar A_Q$, and
if $j\notin S$ then $\bar c_{Qij}=0$ and $(Qi,Qj)\notin \bar A_Q$ - a contradiction.

Assume that the problem is integer-valued. It is straightforward to check that sending one
unit of flow along a minimal path in $(N,\bar A)$ from $s$ to $t$
preserves antisymmetry~\eqref{eq:antisymmetry}, capacity~\eqref{eq:capacity}
and flow conservation~\eqref{eq:conservation} constraints.
We now prove that if $P$ is a minimal path in $(N,\bar A)$ whose endpoints belong to $V$
then sending one unit of flow along $P$ preserves 
base polyhedron constraints~\eqref{eq:base}. Note, $P$ is {\em not} an augmenting path: it does not go from $s$ to $t$.
However, the operation of sending flow along $P$ and the minimality of $P$ are still well-defined.

We use induction on the length of $P$. If $P$ is empty then the claim is trivial.
Suppose $P$ is not empty; since $P$ is minimal and $\bar A_Q$ are transitive,
$P$ must have the form $P=P_1P_2$ where 
$P_1=((i,Qi),(Qi,Qj),(Qj,j))$ 
and $i,j$ are distinct nodes in $Q\in\calQ$. 
Since $(Qi,Qj)\in\bar A_Q$,
sending one unit of flow along $P_1$ preserves base polyhedron constraints.
We prove below that after sending this flow $P_2$ remains a minimal path
in $(N,\bar A)$; the claim will then follow by the induction hypothesis.

Clearly, we need to consider only arcs in $\bar A_Q$ - subsets $\bar A_{Q'}$ for $Q'\in\calQ-\{Q\}$
are not affected. 
Let us denote $\hat f_Q$ to be the residual function after sending the flow
and $\hat A_Q$ to be the corresponding set of arcs.
We have $\hat f_Q(S)=f_Q(S)-[i\in S]+[j\notin S]$ for $S\subseteq Q$, where $[\cdot]$ is the Iverson bracket:
it is 1 if its argument is true, and 0 otherwise. 
We need to show two facts:
\begin{itemize}
\item[(a)] if $(Qk,Ql)\in P_2$ then
$(Qk,Ql)$ remains in $\hat A_Q$; 
\item[(b)] if $(Qk,Ql)$
and $(Qk',Ql')$ are two distinct arcs in $P_2$ occuring in this order
then arc $(Qk,Ql')$ still does not belong to $\hat A_Q$.
\end{itemize}

\myparagraph{Proof of claim (a)} For a set $S\subseteq Q$ denote $[S]=([i\in S],[j\in S],[k\in S],[l\in S])$. \linebreak
If the claim is false then there exists $S$ with $[S]=(?,?,1,0)$ and $\hat f_Q(S)=0$.
Since $(Qk,\!Ql)\!\in\! \bar A_Q$ before sending the flow, we must have 
$\bar f_Q(S)\!=\!\hat f_Q(S)+[i\in S]-[j\notin S]\!>\!0$,
therefore $[S]=(1,0,1,0)$ and $\bar f_Q(S)=1$. By minimality of $P$ arc $(Qi,Ql)$ was not in $\bar A_Q$ before sending the flow,
therefore there exists another set $S'$ with $[S']=(1,?,?,0)$ and $\bar f_Q(S')=0$.
Since $(Qi,Qj),(Qk,Ql)\in\bar A_Q$ we must have $[S']=(1,1,0,0)$.

By submodularity $\bar f_Q(S\cap S')+\bar f_Q(S\cup S')\le \bar f_Q(S)+\bar f_Q(S')=1$,
so one of the set $S\cap S'$, $S\cup S'$ is a minimizer of $\bar f$.
We have $[S\cap S']=(1,0,0,0)$ and $[S\cup S']=(1,1,1,0)$, so either $(Qi,Qj)\notin \bar A_Q$ or
$(Qk,Ql)\notin \bar A_Q$ - a contradiction.

\myparagraph{Proof of claim (b)} For a set $S\subseteq Q$ denote
$[S]=([i\in S],[j\in S],[k\in S],[l\in S],$ $[k'\in S],[l'\in S])$.
Arcs $(Qi,Ql')$ and $(Qk,Ql')$ are not in $\bar A_Q$ before sending the flow,
therefore there exist sets $S$ and $S'$ with $[S]=(1,?,?,?,?,0)$, $[S]=(?,?,1,?,?,0)$
and $\bar f_Q(S)=\bar f_Q(S')=0$. We have $(Qi,Qj),(Qk,Ql),(Qk',Ql')\in \bar A_Q$, therefore
$[S]=(1,1,?,?,0,0)$, $[S']=(?,?,1,1,0,0)$. 

Consider set $S''=S\cup S'$ with $[S'']=(1,1,1,1,0,0)$.
Sets $S$ and $S'$ are minimizers of a submodular function $\bar f$, and thus so is $S''$.
We have $\hat f_Q(S'')=\bar f_Q(S'')-[i\in S'']+[j\notin S'']=0-1+1=0$, which implies the claim.


\section*{Appendix B: proof of proposition~\ref{prop:alpha}}
Let $\varphi^\circ$ be the input flow to the $\Delta$-phase, $S$ to be the set of nodes in $V$ reachable
from $s$ in $(N,\bar A^{2\Delta})$ 
and $\varphi=\AdjustFlow(\varphi^\circ)$. Let $\bar c_{si},\bar c_{it}$ and $\bar f^\Delta_Q$
be residual capacities and functions with respect to flow $\varphi$, and $\bar c^\circ_{si},\bar c^\circ_{it}$
be residual capacities with respect to flow $\varphi^\circ$.
When the previous $2\Delta$-phase terminated, there were no augmenting paths from $s$ to $t$ in $(N,\bar A^{2\Delta})$,
hence $\bar A^{2\Delta}$ cannot have arcs $(i,t)$ for $i\in S$ and $(s,i)$ for $i\in V-S$.
Therefore, $\bar c^\circ_{ti}\le \TwoDeltaCeil-1$ for $i\in S$ and $\bar c^\circ_{is}\le \TwoDeltaCeil-1$ for $i\in V-S$.
Define
\begin{eqnarray}
f^\Delta(S)&=&value(\varphi)+\sum_{i\in S} \bar c_{it} + \sum_{i\in V-S}\bar c_{si} + \sum_{Q\in\calQ} \bar f^\Delta_Q(S\cap Q)
\label{eq:fDelta'}
\end{eqnarray}
Each augmentation in the $\Delta$-phase preserves this equality (assuming that  $\bar c_{si},\bar c_{it}$ and $\bar f^\Delta_Q$ are updated accordingly).
All residual values stay non-negative, therefore the number of augmentations cannot exceed $\left(f^\Delta(S)-value(\varphi)\right)/\DeltaCeil$.
Using~\eqref{eq:fDelta'} and the definition of step S0, we can write
\begin{eqnarray*}
f^\Delta(S)-value(\varphi)&\le& \sum_{i\in S} \bar c^\circ_{it} + \sum_{i\in V-S}\bar c^\circ_{si} 
+ \sum_{Q\in\calQ} \left[ \bar f^\Delta_Q(S\cap Q) + \sum_{i\in Q}|\varphi_{Qi}-\varphi^\circ_{Qi}| \right] \\
& \le & n\cdot(\TwoDeltaCeil-1) + \sum_{Q\in\calQ} \alpha_Q\cdot\DeltaCeil \le \left(2n+\sum_{Q\in\calQ} \alpha_Q\right)\cdot\DeltaCeil
\end{eqnarray*}


\section*{Appendix C: Proof of proposition~\ref{prop:cardinality}}
\myparagraph{Proof of part (a)}
Let $i,i',i''$ be distinct nodes in $Q$ and $(Qi,Qi'),(Qi',Qi'')\in \bar A^\Delta_Q$. 
If \linebreak $z_{i}<z_{i''}$ then obviously $(Qi,Qi'')\in \bar A^\Delta_Q$. Suppose $z_{i}\ge z_{i''}$;
in order to show $(Qi,Qi'')\in \bar A^\Delta_Q$, we need to prove
that
$\min_{k\in \setZ[L(i),R(i'')-1]} \bar g(k) \ge 3\Delta/2$.
Value $z_{i'}$ falls in one of the three intervals $[z_{i},+\infty)$, $(-\infty, z_{i''}]$, $(z_{i''},z_{i})$. These
three cases are considered below.
\begin{itemize}
\item $z_{i'}\ge z_{i}\ge z_{i''}$.
Since arc $(Qi',Qi'')$ belongs $\bar A^\Delta_Q$ and $z_{i'}\ge z_{i''}$,
we must have
$$\min_{k\in \setZ[L(i'),R(i'')-1]} \bar g(k) \ge 3\Delta/2$$
The claim then follows from the fact that $L(i)\ge L(i')$ and so $\setZ[L(i),R(i'')-1]\subseteq\setZ[L(i'),R(i'')-1]$.
\item
$z_{i}  \ge z_{i''}\ge z_{i'}$.
Since arc $(Qi,Qi')$ belongs to $\bar A^\Delta_Q$ and $z_{i}\ge z_{i'}$,
we must have 
$$\min_{k\in \setZ[L(i),R(i')-1]} \bar g(k) \ge 3\Delta/2$$
The claim then follows from the fact that $R(i')\ge R(i'')$ and so $\setZ[L(i),R(i'')-1]\subseteq\setZ[L(i),R(i')-1]$.
\item
$z_{i} > z_{i'} >  z_{i''}$.
We must have
$$\min_{k\in \setZ[L(i),R(i')-1]} \bar g(k) \ge 3\Delta/2
\qquad
\min_{k\in \setZ[L(i'),R(i'')-1]} \bar g(k) \ge 3\Delta/2$$
The claim the follows from the fact that
$R(i') \ge L(i')$ so $\setZ[L(i),R(i')-1] \cup \setZ[L(i'),R(i'')-1]=\setZ[L(i),R(i'')-1]$.
\end{itemize}


\myparagraph{Proof of part (b)} 
The transitivity of $\bar A^\Delta_Q$ and minimality of $P$ implies
that if $(Qi,Qj)\in P$ then the previous and the next arcs of $P$ are respectively $(i,Qi)$ and $(Qj,j)$.
The triple of consecutive arcs $(i,Qi),(Qi,Qj),(Qj,j)$
will be denoted as $(i,j)$, and we will refer to it also as an ``arc''.
Let $P_Q=(i_1,j_1),\ldots, (i_{{d}},j_{{d}})$ be the sequence of all such arcs of $P$
(given in the order that they appear in $P$). Due to the minimality of $P$ all $2d$ nodes involved must be distinct.
It suffices to prove the proposition in the
case when $z_i\ge z_j$ for all arcs $(i,j)$ in this sequence. Indeed,
if there are arcs $(i,j)$ with $z_i < z_j$
then we can push flow through them afterwards - as discussed in section \ref{sec:cardinality},
this cannot violate the base polyhedron constraint.

We thus assume that $z_i\ge z_j$ for arcs $(i,j)\in P_Q$.
Let $(i,j)$ and $(i',j')$ be two consecutive arcs in the sequence.
We claim that $z_j>z_{i'}$. Indeed,
since path $P$ is minimal, arc $(Qi,Qj')$ is not in $\bar A^\Delta_Q$.
If $z_{i'}>z_j$ then $(Qj,Qi')\in \bar A^\Delta_Q$, so by transitivity
we have $(Qi,Qj')\in \bar A^\Delta_Q$ - contradiction.
If $z_{i'}=z_j$ then $(Qi,Qi')\in \bar A^\Delta_Q$ (since $(Qi,Qj)\in \bar A^\Delta_Q$ and $z_{i'}=z_j$),
so by transitivity
we have $(Qi,Qj')\in \bar A^\Delta_Q$ - contradiction.

We showed that $z_{i_1}\ge z_{j_1} > \ldots > z_{i_d} \ge z_{j_d}$.
This implies that $L(i_1)<R(j_1)<\ldots<L(i_d)<R(j_d)$.
Now consider $k\in\setZ[0,m]$; we need to show that $\hat g(k)=g(k)-\sum_{k'=1}^k \hat z^k\ge 0$
where $\hat z$ is the vector after sending $\DeltaCeil$ units of flow through $P$
and $\hat g(\cdot)$ is the corresponding residual function.

\def\A{{S}}
\def\B{{T}}

It follows from the definition of $z^k$ that $Q$ can be partitioned
into two disjoint subsets $\A$, $\B$ with $k$ and $m-k$ nodes,
respectively, such that $z_i\ge z^k\ge z_j$ for any $i\in \A$, $j\in \B$.
Let us introduce the following terminology.
Arc $(i,j)$ in $P_Q$ will be called {\em left-exterior} if
$z_i\ge z_j \ge z^k+\DeltaCeil$ (and thus $i,j\in \A$),
and {\em right-exterior} if $z^k-\DeltaCeil \ge z_i \ge z_j$ (and thus $i,j\in \B$).
Clearly, after the update we have $\hat z_i>\hat z_j\ge z^k$ for left-exterior arcs $(i,j)$
and $z^k\ge \hat z_i>\hat z_j$ for right-exterior arcs $(i,j)$.
An arc in $P_Q$ is called {\em exterior} if it is either left-exterior or right-exterior,
and {\em interior} otherwise.
Note that an interior arc $(i,j)$ must satisfy $z_i\ge z^k \ge z_j$,
which is equivalent to the condition $k\in \setZ[L(i),R(j)]$.
This implies that $P_Q$ can have at most one interior arc.

We now consider three possible cases.
\begin{itemize}
\item All arcs in $P_Q$ are exterior.
Then after the update we have $\hat z_i \ge z^k \ge \hat z_j$
for any $i\in\A$, $j\in \B$,
so $\A$ contains $k$ nodes $i$ with the largest values of $\hat z_i$.
This implies that $\hat g(k)=g(k)-\hat z(\A)$.
Since each arc $(i,j)$ in $P_Q$ either has
both endpoints in $\A$ or both endpoints in $\B$,
we have $\hat z(\A)= z(\A)$,
so $\hat g(k)=\bar g(k)\ge 0$.
\item $P_Q$ has an interior arc $(u,v)$ with $k\in \setZ[L(u),R(v)-1]$;
thus, $\bar g(k)\ge 3\Delta/2$ since $(Qu,Qv)\in \bar A^\Delta_Q$.
We can assume without loss of generality that $u\in \A$ and $v\in \B$.
(Sets $\A$ and $\B$ could have been chosen in this way since $L(u)\le k$ and $R(v)>k$).
After the update we have $\hat z_i \ge z^k \ge \hat z_j$
for any $i\in \A$, $j\in \B$,
so $\A$ contains $k$ nodes $i$ with the largest values of $\hat z_i$.
This implies that $\hat g(k)=g(k)-\hat z(\A)$.
Arc $(u,v)$ is the only one in the sequence which
has exactly one endpoint (namely $u$) in $\A$.
Therefore, $\hat z(\A)= z(\A)+\DeltaCeil$
(where ``$+\DeltaCeil$'' term comes from the update $\hat z_u=z_u+\DeltaCeil$),
so $\hat g(k)=\bar g(k)-\DeltaCeil\ge 3\Delta/2-\DeltaCeil\ge 0$.
\item $P_Q$ has an interior arc $(u,v)$ with $R(v)=k$.
We must have $u,v\in \A$ and $z_u\ge z_v=z^k$.
After the update we have $\hat z_v=z^k-\DeltaCeil$ and 
$\hat z_i \ge z^k \ge \hat z_j$ for any $i\in \A-\{v\}$, $j\in \B$.
Let $(u',v')$ be the arc in $P_Q$
that immediately follows $(u,v)$;
if $(u,v)$ is the last arc in $P_Q$ then we say that $(u',v')$
does not exist.
Two cases are possible:
\begin{itemize}
\item Arc $(u',v')$ does not exist
or $z^k-2\DeltaCeil \ge z_{u'}$. Then $\hat z_v=z^k - \DeltaCeil\ge z_j$
for any $j\in\B$. Thus, $\A$ contains $k$ nodes $i$ with the largest values of $\hat z_i$.
This implies that $\hat g(k)=g(k)-\hat z(\A)$.
Since each arc $(i,j)$ in $P_Q$ either has
both endpoints in $\A$ or both endpoints in $\B$,
we have $\hat z(\A)= z(\A)$,
so $\hat g(k)=\bar g(k)\ge 0$.
\item Arc $(u',v')$ exists and
$z_{u'}=z^k-\DeltaCeil$; thus, $L(u')=R(v)+1=k+1$, $z^{k+1}=z^k-\DeltaCeil$.
After the update $\hat z_v=z^k-\DeltaCeil$, $\hat z_{u'}=z^k$, so the set  $\A'= (  \A -  \{v\}  ) \cup   \{u'\}$
contains $k$ nodes $i$ with the largest values of $\hat z_i$.
This implies that $\hat g(k)=g(k)-\hat z(\A')$.
We have $\hat z(\A')=\hat z(\A) + [\hat z_{u'}-\hat z_v] =  z(\A) + \DeltaCeil $,
so $\hat g(k)=\bar g(k)-\DeltaCeil$.
We now need to show that $\bar g(k)\ge \DeltaCeil$.

Conditions $(Qu,Qv),(Qu',Qv')\in\bar A^\Delta_Q$
imply that $\bar g(k-1)\ge \OnePointFiveDeltaCeil$ and $\bar g(k+1)\ge \OnePointFiveDeltaCeil$. We can write
$$
\begin{array}{rcl}
\bar g(k)  -\bar g(k-1)&=&  [g(k)-g(k-1)] - z^k \\
\bar g(k+1)-\bar g(k)  &=&  [g(k+1)-g(k)] - z^{k+1} \\
\end{array}
$$
Since $g(\cdot)$ is concave, we have $g(k)-g(k-1)\ge g(k+1)-g(k)$. Thus,
$$
\bar g(k)-\bar g(k-1)+z^k\ge  \bar g(k+1)-\bar g(k) + z^{k+1}
$$
\begin{eqnarray*}
2\bar g(k) & \ge & \bar g(k-1)+\bar g(k+1)-[z^k - z^{k+1}] \\
           & \ge & \OnePointFiveDeltaCeil + \OnePointFiveDeltaCeil - \DeltaCeil\ge 3\DeltaCeil-\DeltaCeil=2\DeltaCeil
\end{eqnarray*}
\end{itemize}
\end{itemize}


\myparagraph{Proof of part (c)} 
Let $S$ be the set used in eq.~\eqref{eq:alpha}, and denote $T=Q-S$, $k=|S|$.
We assume that $S\ne \varnothing$ and $S\ne Q$, otherwise the LHS in eq.~\eqref{eq:alpha} would be 0.
Let $i$ be a node in $S$ with the minimum value of $z_i$
and $j$ be a node in $T$ with the maximum value of $z_j$.
Since there was no augmenting path upon termination of the previous $2\Delta$-phase,
set $\bar A^{2\Delta}_Q$ cannot have arc $(Qi,Qj)$.
Therefore, $z_i\ge z_j$ and
$\bar g(\bar k)\le \lfloor 3\Delta \rfloor$ for
some $\bar k\in\setZ[L(i),R(j)-1]$.
The choice of $i,j$ and condition $z_i\ge z_j$ imply that
$\min\{z_{i'}\:|\:z_{i'}\in S\}\ge\max\{z_{j'}\:|\:z_{j'}\in T\}$,
hence $\bar f_Q(S)=\bar g(|S|)=\bar g(k)$.
Thus, we need to show that $\bar g(k)\le \alpha_Q\cdot\DeltaCeil$ 
where $\alpha_Q=3(m-1)$.
If $\bar k=k$ then the claim is obvious. Suppose that $\bar k\ne k$. Two cases are possible:
\begin{itemize}
\item $\bar k>k$. We have $k+1<\bar k+1\le R(j)$, so $z^{k+1}\ge z^{R(j)}=z_j$.
We cannot have $z^{k+1} > z_j$ since in this case there would be at least $k+1$ nodes $j'\in Q$
with $z_{j'}>z_j$; by the choice of $j$ these nodes would belong to $S$,
so we would have $|S|\ge k+1$ - contradiction.
Thus, we must have $z^{k+1}=z^{k+2}=\ldots=z^{R(j)}$.
This implies that function $p(k')=\sum_{k''=1}^{k'} z^{k''}$
is linear in $\setZ[k,R(j)]$.
We have $\bar g(k')=g(k')-p(k')$ where $g(\cdot)$ is a concave function,
therefore $\bar g(\cdot)$ is also concave in $\setZ[k,R(j)]$. There holds $\bar k\in \setZ[k+1,R(j)-1]$,
thus
$$
\bar g(\bar k)\ge
\frac{R(j)-\bar k}{  R(j)-k  }\cdot \bar g(k)
+
\frac{\bar k-k}{   R(j)-k   }\cdot \bar g(R(j))
$$
We have
$\bar g(\bar k)\le \lfloor 3\Delta\rfloor$ and $\bar g(R(j))\ge 0$, therefore
$\bar g(k)\le \frac{R(j)-k}{\stackrel{~}{R(j)-\bar k}} \lfloor 3\Delta\rfloor
\le (m-1)\cdot\lfloor 3\Delta\rfloor\le \alpha_Q\cdot \DeltaCeil$.
\item $\bar k<k$. We have $L(i)\le \bar k<k$, so $z_i=z^{L(i)}\ge z^k$.
We cannot have $z_i > z^k$ since in this case there would be at least $m-k+1$ nodes $i'\in Q$
with $z_{i}>z_{i'}$; by the choice of $i$ these nodes would belong to $T$,
so we would have $|T|\ge m-k+1$ - contradiction.
Thus, we must have $z^{L(i)}=\ldots =z^{k-1}=z^k$.
This implies that function $p(k')=\sum_{k''=1}^{k'} z^{k''}$
is linear in $\setZ[L(i)-1,k]$.
We have $\bar g(k')=g(k')-p(k')$ where $g(\cdot)$ is a concave function,
therefore $\bar g(\cdot)$ is also concave in $\setZ[L(i)-1,k]$. There holds $\bar k\in \setZ[L(i),k-1]$, 
 $\bar g(\bar k)\le \lfloor 3\Delta\rfloor$ and $\bar g(L(i)-1)\ge 0$, so similar to the previous case
we conclude that
$\bar g(k)\le \frac{k-(L(j)-1)}{\stackrel{~}{\bar k-(L(j)-1)}} \lfloor 3\Delta\rfloor
\le (m-1)\cdot\lfloor 3\Delta\rfloor\le \alpha_Q\cdot \DeltaCeil$.
\end{itemize}



\section*{Appendix D: Implementation of $\GetNeighbors^\Delta_Q(Qi)$ for cardinality-dependent terms}
\myparagraph{Case 1} Assume that $f_Q(S)=g(|S|)$ for $S\subseteq Q$ where $g$ is concave.
We use the same notation as in section~\ref{sec:cardinality}.

First, let us describe the data structure for $Q$.
Nodes $i\in Q$ will be grouped into ``supernodes'' according to their value of $z_i$.
The set of supernodes is denoted as $\widetilde Q$.
The cardinality of $\widetilde Q$ equals the number of unique values in the set $\{z_i\:|\:i\in Q\}$.
At each supernode $u\in \widetilde Q$ we store values $z_u=z_i$, $L(u)=L(i)$, $R(u)=R(i)$
where $i$ is a node contained in $u$. We treat supernode $u$ as the set $u=\{Qi\:|\:i\in Q,z_i=z_u\}$.
Supernodes $u$ sorted by their values of $z_u$ will be stored in a doubly-linked list.
Each $u\in \widetilde Q$ also have a pointer to a doubly-linked list of nodes in $u$,
and each node $Qi\in N$ will have a pointer to $u\in \widetilde Q$ with $Qi\in u$.
Finally, we maintain residual function $\bar g(\cdot)$ as an array of size $O(m)$.
It is easy to see that after each augmentation this data structure can be dynamically updated
in $O(m)$ time.

For each supernode $u$ we maintain flag
$\REACHED(u)=\bigwedge_{i\in u}\REACHED(Qi)$; at the beginning of the BFS it is set to $\false$.
Procedure $\GetNeighbors^\Delta_Q(Qi)$ is defined as follows:

~~\underline{$\GetNeighbors^\Delta_Q(Qi)$}
\begin{itemize}
\item Set $\REACHED(Qi):=\true$ and $S:=\varnothing$. Determine supernode $u$ with $Qi\in u$.
\item If $\REACHED(u)$ is {\tt true} then stop, otherwise set $\REACHED(u):=\true$ and continue.
\item If $\min_{k\in \setZ[L(u),R(u)-1]} \bar g(k) \ge 3\Delta/2$ call $\Add(u)$.
\item If $u$ has left neighbor $u^-$ with $z_{u^-}>z_u$ call $\Add(u^-)$ and $\ProcessLeft(u^-)$.
\item If $u$ has right neighbor $u^+$ with $z_u>z_{u^+}$ and $\min_{k\in \setZ[L(u),R(u^+)-1]} \bar g(k) \ge 3\Delta/2$ call
$\Add(u^+)$ and $\ProcessRight(u^+)$.
\end{itemize}

~~\underline{$\ProcessLeft(u)$}
\begin{itemize}
\item If $\REACHED(u)$ is {\tt true} then stop, otherwise set $\REACHED(u):=\true$ and continue.
\item If $u$ has left neighbor $u^-$ with $z_{u^-}>z_u$ call $\Add(u^-)$ and $\ProcessLeft(u^-)$.
\end{itemize}

~~\underline{$\ProcessRight(u)$}
\begin{itemize}
\item If $\REACHED(u)$ is {\tt true} then stop, otherwise set $\REACHED(u):=\true$ and continue.
\item If $u$ has right neighbor $u^+$ with $z_u>z_{u^+}$ and $\min_{k\in \setZ[L(u),R(u^+)-1]} \bar g(k) \ge 3\Delta/2$
call $\Add(u^+)$ and $\ProcessRight(u^+)$.
\end{itemize}

~~\underline{$\Add(u)$}
\begin{itemize}
\item For each node $Qi\in u$ with $\REACHED(Qi)=\false$ set $\REACHED(Qi):=\true$ and add $Qi$ to $S$.
\end{itemize}

The correctness of this procedure should be clear. Note, $\ProcessLeft(u)$ and \linebreak $\ProcessRight(u)$ are only called when
some node $Qi\in u$ has been reached by BFS. If $\REACHED(u)$ is true then all nodes that can be reached from $Qi$
(and from other nodes in $u$) via arcs in $\bar A^\Delta_Q$ have already been added, which justifies statement
``If $\REACHED(u)$ is {\tt true} then stop''.
Steps following this statement will be executed at most once for each supernode $u$,
therefore each node, supernode and element of array $\bar g(\cdot)$ is accessed at most constant
number of times during a single BFS search. Thus, $\beta_Q=O(m)$.

\myparagraph{Case 2} We now assume that $f_Q(S)=g(|S\cap Q'|,|S\cap Q''|)$ for $S\subseteq Q$ where $Q',Q''$ are disjoint
subsets of $Q$. Without loss of generality we can assume that $Q=Q'\cup Q''$. Denote $m'=|Q'|$, $m''=|Q''|$, $m=|Q|=m'+m''$.
Let $y\in\mathbb R^{Q'}$ and $z\in\mathbb R^{Q''}$ be vectors with $y_i=\varphi_{Qi}$ for $i\in Q'$ and $z_j=\varphi_{Qj}$ for $i\in Q''$. We define sequences $(y^1,\ldots,y^{m'})$ and $(z^1,\ldots,z^{m''})$
similar to the case above; $y^k$ and $z^k$ are the $k$-th largest numbers among values $y_i$ and $z_i$, respectively.
Indexes $L(i)$ and $R(i)$ are also defined as in section~\ref{sec:cardinality};
we have $1\le L(i)\le R(i)\le m'$ for $i\in Q'$ and $1\le L(j)\le R(j)\le m''$ for $j\in Q''$.
Let $\bar g(k',k'')=\min\{\bar f_Q(S)\:|\:\:\:|S\cap Q'|=k',|S\cap Q''|=k''\}$ be the ``residual function''.
We have
$$
\bar g(k',k'')=g(k',k'')-\sum_{k=1}^{k'} y^k - \sum_{k=1}^{k''} z^k
$$
It can be seen that $\bar g(\cdot,\cdot)$ is a {\em Monge matrix}~\cite{Burkard:DAM96}, i.e.\ for any 
$0\le k'_1<k'_2\le m'$ and $0\le k''_1<k''_2\le m''$ there holds 
$\bar g(k'_1,k''_1)+\bar g(k'_2,k''_2)\le \bar g(k'_1,k''_2)+\bar g(k'_2,k''_1)$.
This follows from
$\bar f_Q(S'\cap S'')+\bar f_Q(S'\cup S'')\le \bar f_Q(S')+\bar f_Q(S'')$
where $S'$ contains first $k'_1$ nodes of $Q'$ and first $k''_2$ nodes of $Q''$,
and $S''$ contains first $k'_2$ nodes of $Q'$ and first $k''_1$ nodes of $Q''$.
(We assume that nodes in $Q'$ and $Q''$ are sorted so that components $y_i$ and $z_i$ are non-increasing.)
For a row $k'\in\setZ[0,m']$ let $k''(k')\in\setZ[0,m'']$ be the column that contains
the leftmost minimum entry in row $k'$. Thus, 
$k''(k')=\min\{k''\in\setZ[0,m'']\:|\:\bar g(k',k'')=\min_{k''} \bar g(k',k'')\}$.
It is known~\cite{Burkard:DAM96} that Monge matrices are {\em monotone},
i.e.\ $k''(0)\le k''(1)\le \ldots k''(m')$. Furthermore, they are {\em totally monotone},
i.e.\ every submatrix is monotone. As shown by Aggarwal et al.~\cite{Aggarwal:87},
indexes $k''(0),\ldots,k''(m')$ for a totally monotone matrix can be computed in $O(m)$ time.

We can describe data structures for implementing $\GetNeighbors^\Delta_Q(Qi)$.
Nodes in $i\in Q'$ will be grouped into supernodes according to the values $y_i$
analogously to case 1. A similar data structure will be used for nodes in $Q''$.
We will maintain an array of cumulative sums $\sum_{k=1}^{k'}y^k$ for $k'\in\setZ[0,m']$
and $\sum_{k=1}^{k''}z^k$ for $k''\in\setZ[0,m'']$, which will allow computing $\bar g(k',k'')$ in $O(1)$ time.
At the beginning of each BFS we will compute indexes $k''(k')$ for $k'\in\setZ[0,m']$ using
the algorithm in~\cite{Aggarwal:87} and also indexes 
$k'(k'')=\min\{k'\in\setZ[0,m']\:|\:\bar g(k',k'')=\min_{k'} \bar g(k',k'')\}$
for each column $k''\in\setZ[0,m'']$.

Arcs in $\bar A^\Delta_Q$ can be split into four groups $\bar A_{00}, \bar A_{01}, \bar A_{10}, \bar A_{11}$
where $\bar A_{\alpha\beta}=\{(Qi,Qj)\in\bar A^\Delta_Q\:|\:[i\in Q']=\alpha,[j\in Q'']=\beta\}$
and $[\cdot]$ is the Iverson bracket: it is 1 if its argument is true, and 0 otherwise.
Consider the version of $\GetNeighbors^\Delta_Q(Qi)$ that processes only arcs in a specific
group, rather than all arcs in $\bar A^\Delta_Q$. It suffices to show how to implement
such procedure for each of the four groups; these procedures will be called sequentially.

First, consider arcs in $\bar A_{11}$. 
Using the same argumentation as in section~\ref{sec:cardinality}
we conclude that sending flow from a node $Qi$ to another node $Qj$ ($i,j\in Q'$)
is possible if and only if one of the two conditions hold:
(a) $y_i<y_j$; (b) $y_i\ge y_j$ and $\min_{k'\in \setZ[L(i),R(j)-1]}\bar g'(k')\ge 1$
where we defined $\bar g'(k')=\min_{k''} \bar g(k',k'')$.
Thus, the set $\bar A_{00}$ is constructed completely analogously
to the set $\bar A^\Delta_Q$ in section~\ref{sec:cardinality}
except that function $\bar g$ is replaced with $\bar g'$ and
threshold $3\Delta/2$ is replaced with 1. Accordingly,
we can use an obvious adaptation of the procedure for the case 1.
Note, $\bar g'(k')$ can be evaluated in $O(1)$
time using arrays of indexes $k''(k')$ and cumulative sums for vectors $y$ and $z$.
Arcs in $\bar A_{00}$ can be handled in a similar way. It remains to show how to handle
arcs in $\bar A_{10}$ (the set $\bar A_{01}$ will follow by symmetry).

Consider nodes $i\in Q'$, $j\in Q''$. Sending $\DeltaCeil$ units of flow from $i$ to $j$ via $Q$,
i.e.\ the operation $y_i:=y_i+\DeltaCeil$, $z_j:=z_j-\DeltaCeil$, affects function $\bar g(\cdot,\cdot)$
as follows: values $\bar g(k',k'')$ for $(k',k'')\in\setZ[L(i),m']\times\setZ[1,R(j)-1]$
are decreased by $\DeltaCeil$ and values
$\bar g(k',k'')$ for $(k',k'')\in\setZ[1,L(i)-1]\times\setZ[R(j),m'']$
are increased by $\DeltaCeil$. Thus, sending flow is possible if and only if
\begin{equation*}
\min\left\{\bar g(k',k'') \: | \: (k',k'') \in K(L(i),R(j)) \right\} > 0 \;\;,
\quad K(a,b)=\setZ[a,m']\times\setZ[0,b-1]
\end{equation*}
There holds $K(a,b_1)\subset K(a,b_2)$ for $b_1<b_2$, therefore
the set of arcs in $\bar A_{10}$ from $Qi$ have the form
$\{(Qi,Qj)\:|\:R(j)\le b(L(i))\}$ where 
\begin{equation}
b(a)=\max\left\{b\in \setZ[1,m'']\:\mbox{\Big |}\:\min_{(k',k'')\in K(a,b)}\bar g(k',k'')>0\right\}
\qquad a\in\setZ[1,m']
\label{eq:dgasdg}
\end{equation}
(If the set in~\eqref{eq:dgasdg} is empty then we assume that $b(a)=0$.)
We compute indexes $b(a)$ at the beginning of BFS in linear time
using the following recursion:
\begin{eqnarray*}
b(m')&=&k''(m') \quad\qquad \mbox{(since $\min_{k''} \bar g(m',k'')=\bar g(m',m'')=0$)} \\
b(a)~&=&
\begin{cases}
b(a+1)                  & \mbox{if } \bar g(a,k''(a))>0 \\
\min\{b(a+1),k''(a)\} & \mbox{if } \bar g(a,k''(a))=0
\end{cases}\qquad\forall a\in\setZ[1,m'-1]
\end{eqnarray*}
Note that $0\le b(1)\le \ldots \le b(m')\le m''$. Procedure $\GetNeighbors^\Delta_Q(Qi)$, $i\in Q'$
for the set of arcs $\bar A_{10}$ is implemented as follows. First, we locate the rightmost 
supernode $v\subseteq Q''$ satisfying $R(v)\le b(L(i))$. (Pointers to these supernodes
for each supernode $u\subseteq Q'$ can be computed at the beginning of BFS.) We then call procedure $\Add(v)$,
which is defined as in the case 1, and procedure
$\ProcessLeft_{10}(v)$ defined as follows:

~~\underline{$\ProcessLeft_{10}(v)$}
\begin{itemize}
\item If $\REACHED(v)$ is {\tt true} then stop, otherwise set $\REACHED(v):=\true$ and continue.
\item If $v$ has left neighbor $v^-$ with $z_{v^-}>z_v$ call $\Add(v^-)$ and $\ProcessLeft_{10}(v^-)$.
\end{itemize}


\small
\bibliographystyle{plain}

\end{document}